# Cross-Learning Fine-Tuning Strategy for Dysarthric Speech Recognition Via CDSD database

*Qing Xiao* [1], *Yingshan Peng*[2], *PeiPei Zhang*[3]

[1] College of Information Science and Engineering, Xinjiang University, Urumqi, China
[2] College of Information Science and Engineering, Xinjiang University, Urumqi, China
[3] College of Information Science and Engineering, Xinjiang University, Urumqi, China

20221401236@stu.xju.edu.cn

## Abstract

Dysarthric speech recognition faces challenges from severity variations and disparities relative to normal speech. Conventional approaches individually fine-tune ASR models pre-trained on normal speech per patient to prevent feature conflicts. Counter-intuitively, experiments reveal that multi-speaker fine-tuning (simultaneously on multiple dysarthric speakers) improves recognition of individual speech patterns. This strategy enhances generalization via broader pathological feature learning, mitigates speaker-specific overfitting, reduces per-patient data dependence, and improves target-speaker accuracy—achieving up to 13.15% lower WER versus single-speaker fine-tuning.

**Index Terms**: Dysarthria, Dysarthric Speech Recognition, Mandarin, Cerebral Palsy, FinTuning

## 1. Introduction

Speech, as the most critical medium for daily information exchange and interpersonal connection, fundamentally underpins human collaboration[1]. Individuals with advanced communication skills demonstrate enhanced cooperative capabilities and achieve greater professional success [2]. However, people with dysarthria — a motor speech disorder impairing articulatory precision — face significant barriers in expressing their thoughts verbally [3]. This impairment often leads to misunderstandings, social isolation, and discrimination, which may contribute to long-term psychological distress due to persistent lack of recognition[4]. Therefore, to empower individuals with dysarthria in expressing their thoughts and establishing social connections, it has been found that the advancement of Dysarthric Speech Recognition (DSR) technology can effectively convert their impaired speech into text, thereby facilitating comprehension among listeners.

With technological advancements, mainstream Automatic Speech Recognition (ASR) systems—such as FireRedASR-AED [5], Seed-ASR [6], and Qwen-Audio [7]—have achieved notable maturity. These models consistently reduce the Character Error Rate (CER) below 3% on benchmark datasets like AISHELL-1 [8], demonstrating highly accurate recognition and translation capabilities for typical speech. Furthermore, the ASR field benefits from extensive data resources: Chinese datasets include AISHELL-1 (178 hours, 400 speakers), AISHELL-2 (1,000 hours, 1,991 speakers) [9], and WenetSpeech (12,400 hours sourced from online videos) [10]. Complementing these, robust English datasets such as LibriSpeech (1,000 hours, ~2,484 speakers) [11] provide equally ample resources for training robust models.

However, a primary challenge in DSR research lies in the acute scarcity of datasets, particularly for Mandarin Chinese—a deficit stemming from two inherent barriers: physiological constraints impairing dysarthric individuals' capacity for prolonged or complex utterances, and demographic limitations where the dysarthric population represents a small subset relative to typical speakers, complicating large-scale recruitment. Although English dysarthria corpora are comparatively mature, their scale remains inadequate: the Torgo corpus (jointly developed by University of Toronto and Blythedale Children's Hospital) [12], targeting cerebral palsy or amyotrophic lateral sclerosis (ALS) patients, contains only 8 speakers each contributing <30 minutes of speech; similarly, UA-Speech (University of Illinois) [13] comprises 15 cerebral palsy speakers but is limited to isolated-word recordings, lacking continuous conversational data critical for modeling natural communication dynamics.

Currently available Chinese dysarthria datasets are limited to the Cantonese Dysarthric Speech Corpus (CUDYS) [14] and the Mandarin Subacute Stroke Dysarthria Multimodal (MSDM) database [15]. The CUDYS primarily investigates acoustic features of dysarthric speech, including articulation, rhythm, pitch, and intensity. Meanwhile, the MSDM recruits participants with subacute stroke, capturing audio via professional-grade microphones while synchronously recording facial motion data during speech production. However, both databases suffer from relatively small datasets—each containing under 10 hours of dysarthric speech—rendering them insufficient for training comprehensive speech recognition models.

Fortunately, the recent release of the Chinese Dysarthria Speech Database (CDSD) [16] by the Chinese Academy of Sciences represents the largest publicly available Mandarin dysarthric speech dataset to date. Based on the CDSD database, their baseline experiments using the WeNet [17] model revealed critical limitations in dysarthric speech recognition: conventional ASR models performed poorly without fine-tuning—evidenced by a severe incompatibility rate of 62.14% when applying the WenetSpeech pre-trained model directly to dysarthric speech—and highlighted two key issues: (1) fine-tuning for dysarthric speech increased the CER on the AISHELL-1 benchmark (normal speech) by 1.5–3.2%, indicating mutual exclusivity between pathological and normative speech features; and (2) while speaker-dependent fine-tuning achieved a CER of 7.37% for individual dysarthric speakers, performance plummeted to 55.54% when tested on a multi-speaker dataset, underscoring substantial acoustic heterogeneity among individuals with dysarthria [18].

Building upon the CDSD database, we selected WeNet—an end-to-end speech recognition toolkit co-developed by Mobvoi and Northwestern Polytechnical University—as the baseline system. Its core innovation leverages the a unified two-pass plus (U2++) framework to unify streaming and non-streaming ASR by synergistically integrating connectionist temporal classification (CTC) and attention mechanisms. The shared encoder, constructed with stacked Transformer or Conformer layers, processes acoustic inputs while deliberately restricting access to right-side context to maintain low-latency equilibrium during streaming inference.

Conventional dysarthric speech recognition (DSR) methodologies predominantly rely on speaker-specific fine-tuning to address pronunciation variations through personalized adaptation. While this approach effectively leverages individual speech patterns, it introduces two critical constraints. Extensive data requirements: Necessitates collecting substantial speech samples (>1 hour per patient), creating significant barriers for severely impaired individuals. Limited cross-speaker generalization: Fails to exploit latent articulatory relationships across dysarthric populations.

This study proposes Cross-Speaker Joint Fine-Tuning, overcoming these constraints through heterogeneous data co-training. The core innovation leverages inter-speaker pronunciation discrepancies as intrinsic data augmentation. This approach demonstrates that aggregating ultra-sparse samples from diverse dysarthric individuals can yield superior generalization capability compared to intensive single-speaker training.

Our empirical analysis of the training efficacy between Part A and Part B offers a new perspective to prior research. While Wang et al. demonstrated "superiority of Part A over Part B" under smaller-scale model conditions—attributed to Module A's inclusion of more speakers—our scaled experiments with larger WeNet models reveal that data diversity ceases to be the primary determinant when dataset duration falls below a critical threshold.

In fine-tuning experiments employing phoneme-based modeling units, we observed suboptimal performance due to the insufficient information capacity of phonemes alone—while phonemes effectively capture pronunciation features, their inherent lack of lexical-semantic information restricts holistic language pattern acquisition. Directly fine-tuning the entire model under such constraints thereby fails to achieve ideal results.

Leveraging the cross-age/etiology/device characteristics of the CDSD corpus, we design three experimental groups to rigorously evaluate the following hypotheses:

H1: Cross-speaker joint fine-tuning achieves superior adaptation efficacy compared to speaker-specific approaches;

H2: Data scaling effects exhibit stronger influence than speaker population size;

H3: The inherent information scarcity in phoneme-based representations (acoustic precision at the cost of semantic context) creates an acoustic-semantic misalignment that naive end-to-end fine-tuning fails to sufficiently exploit, necessitating specialized modeling architectures.

## 2. Dataset

CDSD comprising recordings from 44 individuals with dysarthria divided into Part A and Part B. Part A includes 44 hours of speech data (1 hour per participant from all 44 speakers), while Part B adds 80 hours from 8 participants (10 hours each), totaling 124 hours of dysarthric speech alongside 9 hours of synchronized video from 9 participants. To accommodate varying literacy levels, the text corpus primarily references the AISHELL-1 dataset and speeches by Chinese primary/secondary students. Data collection utilized participants' smartphones and ZOOM F8n multi-track field recorders, allowing fragmented recording sessions in home environments to capture dysarthric speech samples.

The dataset comprises recordings from 44 speakers, with 39 participants aged 18 or above and 5 under 18 years old. Additionally, video recordings were captured for 9 speakers. All participants were informed of their right to withdraw at any time during the process. To protect speaker privacy, personally identifiable names were replaced by anonymized serial numbers. Further details regarding speaker-specific characteristics—including etiology of dysarthria, recording environment, age, accent, and other metadata—are comprehensively presented in TABLE 1.

Table 1. *The overall information of the speakers.*

| Factor | Categorization | Overall(N=44) |
| --- | --- | --- |
| Sex | Femal/Male | 18/26 |
| Age | Adults/Children | 39/5 |
| Etiology | Cerebral Palsy/Other Disease | 33/11 |
| Recording devices | Smartphone/ZOOM F8n | 39/5 |

The annotation of dysarthric speech faced significant challenges due to substantial acoustic variations between individuals and deviations from healthy speech patterns. Issues included reading errors by participants, severe articulation deficits obscuring pronunciation, and strong regional accents, complicating speech recognition for annotators. To address this, five trained annotators employed the AIBIAOKE platform under standardized protocols. After quality checks and speaker re-recording if needed, audio data were imported for verbatim transcription, ensuring precise audio-text alignment. 0.1-second buffers were added around speech segments to prevent clipping, and non-speech sounds ≥0.5s (e.g., noise, laughter, coughing) were labeled as "NOISE".

Developed by Mengyi Sun et al., the Chinese Dysarthria Speech Database (CDSD) comprises dysarthric speech data from 44 dysarthric speakers, organized into two subsets, Part A: Includes 44 hours of audio recordings (1 hour per speaker) from all 44 participants and Part B: Contains an additional 80 hours of recordings (10 hours per speaker) from a subset of 8 speakers. but the data of Speaker #2 in Part B was removed from the experiment because it was incorrectly annotated. So Part B remains 7 speakers for performance comparison, with each contributing roughly ten hours of data, amounting to a total of approximately 70 hours in Table 2.

TABLE 2. *CDSD divided strategy.*

|       | #   | D/P | Total audio |
|-------|-----|-----|-------------|
| PartA | 44  | 1h  | 44h         |
| PartB | 7*  | 10h | 70h         |

a # indicates the amount of the speakers.
b D/P represents the recording duration per speaker.
c *The speakers in partB were also involved in partA.

We modified the DSR modeling units from character-level units to phoneme-level units, performing end-to-end fine-tuning at the phoneme level. Using Pypinyin, we first constructed a phoneme dictionary by converting the training set texts from AISHELL-1 into phoneme sequences. Each phoneme (including tone markers) was stored upon its first occurrence, establishing a one-to-one mapping between IDs and phonemes. Subsequently, the training, validation, and test sets from both AISHELL-1 and CDSD were transformed from character sequences to phoneme sequences, preparing them for subsequent model training. The entire transformation process is visualized in Figure 1.

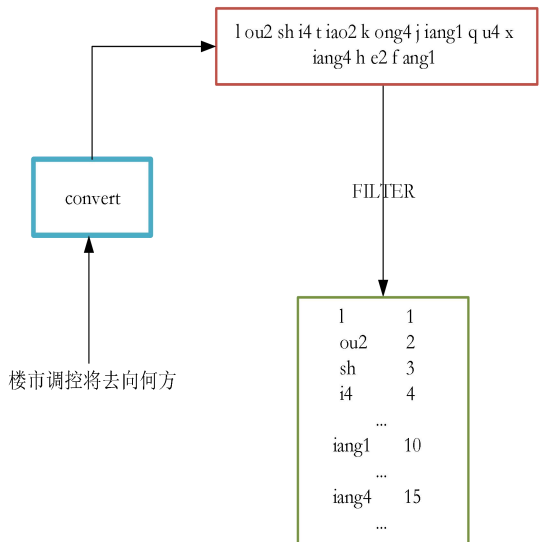

Figure 1: *Phoneme Unit Processing Workflow. Arrows indicate unidirectional text processing. The "Convert" module executes grapheme-to-phoneme (G2P) conversion using Pypinyin, where digits 0-5 denote tone attributes (exclusively recognized as tonal markers when appearing at syllable endings), with 5 indicating neutral tone and 0 representing toneless syllables, compliant with the Style.TONE3 specification. Red segments depict phonemes after grapheme conversion (e.g., "l ou2 sh i4 t iao2 k ong4" for "楼市调控" ), while green segments denote filtered phonemes post-conversion. The "FILTER" module establishes a one-to-one mapping between unique numeric IDs and phonemes upon their initial occurrence for lexical initialization.*

## 3. Experiments

Experiments were conducted on PartB of the Chinese Dysarthria Speech Database (CDSD), with each speaker's samples in PartB analyzed individually. The data is segmented into training, development, and test sets in an 8:1:1 ratio. W10 refers to the 10-hour speech data of a dysarthria speaker in PartB. We initially employed the WeNetSpeech pre-trained model (officially released by WeNet) as the baseline, subsequently fine-tuning it using both: the speaker-specific W10 dataset (10-hour dysarthric speech), and the PartB training and development sets.Evaluation was conducted on the W10 test set and PartB test set, with results comprehensively documented in TABLE 2.

These experimental results demonstrate that the Multi-Speaker Fine-Tuning Strategy significantly enhances personalized adaptation efficacy. In educational psychology, interleaved practice一which involves systematically alternating between distinct knowledge modules一has been empirically shown to yield substantially superior learning outcomes compared to conventional mathematics training using blocked practice. During our experiments, speaker 04 and speaker 06 exhibited increased character error rates (CERs) after PartB and W10 together fine-tuning in Figure 2. This observation suggests that cross-speaker training can improve model generalization capability to some extent, as it compels the model to develop more robust feature representations across heterogeneous speech patterns.

TABLE 2. *Multi-Speaker Fine-Tuning Strategy Results*

| ID | 01 | 04 | 06 | 08 | 09 | 12 | 20 |
|---|---|---|---|---|---|---|---|
| W10 | 8.91 | 32.3 | 41.21 | 23.19 | 24.69 | 12 | 8.2 |
| PartB | 7.47 | 28.76 | 28.06 | 16.83 | 17.34 | 6.77 | 5.06 |

[a] *ID represents specific speakers in partB.*
[b] *W10 refers to the 10-hour speech data of a dysarthria speaker in PartB.*

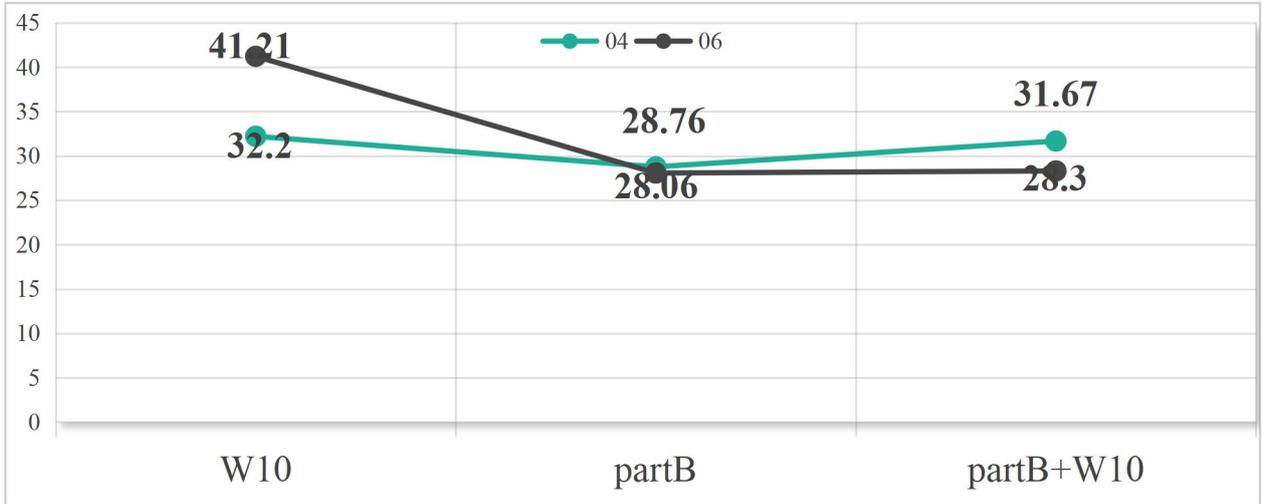

FIG. 2: *the vertical axis values represent Character Error Rates (CERs), while the horizontal axis denotes three experimental conditions: W10 (referring to the 10-hour dysarthria speech subset from PartB), PartB (baseline dataset), and Fine-tuned PartB+W10 (sequential fine-tuning: PartB followed by W10). The green line illustrates performance for Speaker 04, and the black line depicts results for Speaker 06, demonstrating comparative CER outcomes across these training scenarios.*

We designed an experiment targeting speaker W04 due to its slightly divergent trend in TABLE II compared to other speakers. To investigate the impact of PartB composition on W04 fine-tuning efficacy, We compared five configurations: PartB without speaker 20, PartB without speaker 1, PartB without speaker 12, PartB without speakers 1 and 12, and PartB without speakers 20 and 12. As summarized in TABLE III, the lowest CER of 25.17% was achieved when fine-tuning on PartB excluding speaker 20. This result demonstrates that increasing the number of speakers in PartB does not necessarily yield better performance for target speaker adaptation.

TABLE 3. *W04 fine-tune with different configurations.*

| | PartB | PartB-20 | PartB-1 | PartB-12 | PartB-20-12 | PartB-1-12 |
|---|---|---|---|---|---|---|
| W04 | 28.76 | 25.17 | 29.11 | 29.7 | 27.46 | 29.93 |

[a] *The sub "-" indicates PartB without specific speakers.*
[b] *W10 refers to speaker 04 in PartB.*

The scales of PartA/B of CDSD, AISHELL-1, WnetSpeech are 40+/70+, 100+, 10000+ hours,respectively. It is worth noting that the number of parameters differs between the two official WeNet pre-trained models. Specifically, the AISHELL-1 pre-trained model and the model trained directly on dataset PartA/B in our experiments share a same model structure and identical parameter count of 48,347,803. However, the WenetSpeech pre-trained model contains 122,553,574 parameters. Experimental validation reveals that on identical pre-trained models, In TABLE 4, CERs on the CDSD PartB dataset are consistently lower than those on PartA, with the optimal performance consistently achieved using the WenetSpeech pre-trained model. This observation diverges from the findings reported by Yan Wang who noted:"Despite PartB's longer duration (44 hours) compared to PartA (70 hours), PartA encompasses a significantly larger number of subjects (7 vs. 44). This suggests that under subject-quantity constraints, models exhibit poorer generalization capacity, failing to capture diverse speech characteristics." We hypothesize that this discrepancy arises due to differences in model capacity. The Wenet model's has larger architecture so, the advantage of subject diversity becomes pronounced when fine-tune data duration is sufficient.

TABLE 4. *CERs with Fbank on various pre-trained models.*

|       | PartX | AISHELL1→PartX | WenetSpeech→PartX |
|-------|-------|----------------|-------------------|
| PartA | 35.68 | 25.87          | 20.58             |
| PartB | 29.17 | 21.16          | 18.35             |

[a] The arrow "→" indicates that pre-training is conducted on the databases of the former, followed by fine-tuning on the Part A or Part B of the CDSD database.

To validate the effectiveness of direct phoneme-based fine-tuning, we utilized the Pinyin tool to convert all sentences in the AISHELL-1 and CDSD datasets into phoneme sequences. Three distinct approaches were implemented: (1) fine-tuning a character-level AISHELL-1 pre-trained model with character-level PartA/B data; (2) fine-tuning a phoneme-level AISHELL-1 pre-trained model with phoneme-level PartA/B data; and (3) fine-tuning a character-level AISHELL-1 pre-trained model with phoneme-level PartA/B data. Experimental results revealed that when a model trained and fine-tuned on character-level units was applied to its own test-set outputs (originally predicted as Chinese characters), and these character sequences were converted to phoneme sequences using the identical Pinyin tool, the resulting Phoneme Error Rates (PERs) in TABLE 5 was superior to that achieved by the directly phoneme-trained model. Based on the experimental findings and technical analysis, we conclude that character-level representations inherently capture richer semantic and contextual information, which enhances model performance; meanwhile, naive full-model fine-tuning may disrupt pre-trained knowledge alignment and should be replaced with carefully designed layer-specific tuning to preserve foundational acoustic-semantic patterns.

TABLE 5. *Impact of Modeling Unit Transfer Strategies on PERs*

| AISHELL1 | W→W  | P→P   | W→P   |
|----------|------|-------|-------|
| PartA    | 22.66| 32.29 | 25.78 |
| PartB    | 15.07| 21.08 | 17.64 |

[a] The arrow '→' indicates the modeling units: those preceding the arrow represent the pre-trained model, while those following represent the fine-tuned model
[b] P refers phoneme and W refers word.

## 4. Discussion

Based on the experimental findings, we hypothesize that the significant CERs increase observed for Speaker 04 after sequential fine-tuning (PartB → W04) versus direct PartB fine-tuning stems from conflicting acoustic characteristics between Speaker 04 and other speakers in PartB. However, the specific nature of this inter-speaker conflict among dysarthric populations remains unclear. Future work should investigate the distinctive speech attributes that cause training confusion in multi-speaker joint datasets, enabling optimized speaker selection criteria for cross-training. Furthermore, to deepen the insights from Experiment 2, subsequent studies could systematically evaluate how speaker diversity and speech duration interact with model scale (e.g., comparing WeNet versus ESPnet architectures), thereby clarifying the relationship between subjects diviserty and model generalization. For the third experiment exploring phoneme-level modeling units, while finer granularity theoretically facilitates parameter sharing between normative and dysarthric speech models by isolating unaffected phonemes, our end-to-end full-model fine-tuning approach proved suboptimal. Further investigation should identify which model layers (e.g., encoder lower layers for acoustic features vs. decoder layers for semantic context) benefit most from phoneme-specific tuning to maximize cross-population parameter sharing.

## 5. Conclusions

In this study, we conducted three critical experiments: The first experiment validated our proposed multi-speaker cross-training strategy, revealing that for all seven speakers (each with 10 hours of speech data in CDSD), their individual Character Error Rates (CERs) significantly decreased compared to direct fine-tuning using solely their personal speech data. Further analysis paradoxically showed that sequential fine-tuning—first with multi-speaker data followed by speaker-specific refinement—increased CERs for Speakers 04 and 06. Moreover, contrary to conventional wisdom, expanding the speaker population in PartB did not monotonically improve performance. The second experiment supplemented prior research by demonstrating that for larger-scale models, speech duration conditions outweighed speaker quantity conditions during dataset training, establishing duration as a more decisive factor than speaker diversity for model training effectiveness. The third experiment investigated modeling-unit efficacy, contrasting phoneme-based and character-based approaches; results confirmed that direct full-model fine-tuning with phoneme-based units yielded suboptimal performance, underscoring the inherent limitations of phonemic representations in capturing semantic-contextual alignment.

## 5. References

[1] J. H. Manson, G. A. Bryant, M. M. Gervais, and M. A. Kline, "Convergence of speech rate in conversation predicts cooperation," *Evol. Hum. Behav.*, vol. 34, no. 6, pp. 419–426, Nov. 2013, doi: 10.1016/j.evolhumbehav.2013.08.001.
[2] R. E. Riggio, "Assessment of basic social skills.," *J. Pers. Soc. Psychol.*, vol. 51, no. 3, pp. 649–660, Sep. 1986, doi: 10.1037/0022-3514.51.3.649.
[3] S. Braithwaite and J. Holt-Lunstad, "Romantic relationships and mental health," *Curr. Opin. Psychol.*, vol. 13, pp. 120–125, Feb. 2017, doi: 10.1016/j.copsyc.2016.04.001.
[4] A. D. Palmer, J. T. Newsom, and K. S. Rook, "How does difficulty communicating affect the social relationships of older adults? An exploration using data from a national survey," *J. Commun. Disord.*, vol. 62, pp. 131–146, Jul. 2016, doi: 10.1016/j.jcomdis.2016.06.002.
[5] K.-T. Xu, F.-L. Xie, X. Tang, and Y. Hu, "FireRedASR: Open-Source Industrial-Grade Mandarin Speech Recognition Models from Encoder-Decoder to LLM Integration," Jan. 24, 2025, *arXiv*: arXiv:2501.14350. doi: 10.48550/arXiv.2501.14350.
[6] Y. Bai *et al.*, "Seed-ASR: Understanding Diverse Speech and Contexts with LLM-based Speech Recognition," Jul. 10, 2024, *arXiv*: arXiv:2407.04675. doi: 10.48550/arXiv.2407.04675.
[7] Y. Chu *et al.*, "Qwen-Audio: Advancing Universal Audio Understanding via Unified Large-Scale Audio-Language Models," Dec. 21, 2023, *arXiv*: arXiv:2311.07919. doi: 10.48550/arXiv.2311.07919.
[8] H. Bu, J. Du, X. Na, B. Wu, and H. Zheng, "AISHELL-1: An open-source Mandarin speech corpus and a speech recognition baseline," in *2017 20th Conference of the Oriental Chapter of the International Coordinating Committee on Speech Databases and Speech I/O Systems and Assessment (O-COCOSDA)*, Seoul: IEEE, Nov. 2017, pp. 1–5. doi: 10.1109/ICSDA.2017.8384449.
[9] J. Du, X. Na, X. Liu, and H. Bu, "AISHELL-2: Transforming Mandarin ASR Research Into Industrial Scale," Sep. 13, 2018, *arXiv*: arXiv:1808.10583. doi: 10.48550/arXiv.1808.10583.
[10] B. Zhang *et al.*, "WENETSPEECH: A 10000+ Hours Multi-Domain Mandarin Corpus for Speech Recognition," in *ICASSP 2022 - 2022 IEEE International Conference on Acoustics, Speech and Signal Processing (ICASSP)*, Singapore, Singapore: IEEE, May 2022, pp. 6182–6186. doi: 10.1109/ICASSP43922.2022.9746682.
[11] V. Panayotov, G. Chen, D. Povey, and S. Khudanpur, "Librispeech: An ASR corpus based on public domain audio books," in *2015 IEEE International Conference on Acoustics, Speech and Signal Processing (ICASSP)*, South Brisbane, Queensland, Australia: IEEE, Apr. 2015, pp. 5206–5210. doi: 10.1109/ICASSP.2015.7178964.
[12] F. Rudzicz, A. K. Namasivayam, and T. Wolff, "The TORGO database of acoustic and articulatory speech from speakers with dysarthria," *Lang. Resour. Eval.*, vol. 46, no. 4, pp. 523–541, Dec. 2012, doi: 10.1007/s10579-011-9145-0.
[13] Heejin Kim, Mark Hasegawa Johnson, Jonathan Gunderson, Adrienne Perlman, Thomas Huang, Kenneth Watkin, Simone Frame, Harsh Vardhan Sharma, Xi Zhou, "UASpeech", IEEE Dataport, March 17, 2023, doi:10.21227/f9tc-ab45.
[14] K. H. Wong, Y. T. Yeung, E. H. Y. Chan, P. C. M. Wong, G.-A. Levow, and H. Meng, "Development of a Cantonese dysarthric speech corpus," in *Interspeech 2015*, ISCA: ISCA, Sep. 2015, pp. 329–333. doi: 10.21437/interspeech.2015-149.
[15] J. Liu *et al.*, "Audio-video database from subacute stroke patients for dysarthric speech intelligence assessment and preliminary analysis," *Biomed. Signal Process. Control*, vol. 79, p. 104161, Jan. 2023, doi: 10.1016/j.bspc.2022.104161.
[16] Y. Wan *et al.*, "CDSD: Chinese Dysarthria Speech Database," in *Interspeech 2024*, ISCA: ISCA, Sep. 2024, pp. 4109–4113. doi: 10.21437/interspeech.2024-1597.
[17] B. Zhang *et al.*, "WeNet 2.0: More Productive End-to-End Speech Recognition Toolkit," Jul. 05, 2022, *arXiv*: arXiv:2203.15455. doi: 10.48550/arXiv.2203.15455.
[18] Y. Wang *et al.*, "CDSD: Chinese Dysarthria Speech Database," Sep. 2024, pp. 4109–4113. doi: 10.21437/Interspeech.2024-1597.